\documentclass[12pt]{article}

\begin{document}

\title{\normalsize \hfill UWThPh-2000-32 \\[1cm] \LARGE
The seesaw mechanism at arbitrary order: \\
disentangling the small scale from the large scale}

\author{W.\ Grimus \\
\small Universit\"at Wien, Institut f\"ur Theoretische Physik \\
\small Boltzmanngasse 5, A--1090 Wien, Austria
\\[3.6mm]
L.\ Lavoura \\
\small Universidade T\'ecnica de Lisboa \\
\small Centro de F\'\i sica das Interac\c c\~oes Fundamentais \\
\small Instituto Superior T\'ecnico, P--1049-001 Lisboa, Portugal}

\date{5 October 2000}

\maketitle

\begin{abstract}
We develop a recipe which allows one
to recursively and uniquely decouple
the large scale from the small scale
in mass matrices of the
seesaw type, up to
any order in the inverse of the large scale.
Our method allows
one to calculate the mass matrix of the light neutrinos
with arbitrary precision.
The same method
can be applied in the case of quark mass matrices
in an extension of the Standard Model with vector-like quarks
which have mass terms at a scale
much higher than the electroweak scale.
\end{abstract}

\vspace{6mm}

\section{Introduction}

Neutrino physics might prove to be the first window
to physics beyond the
Standard Model. The reason for this is the evidence for neutrino oscillations
found, in particular, in experiments measuring the atmospheric neutrino flux
\cite{review}. Most probably, physics beyond the Standard Model is
associated with a new mass or energy scale above the electroweak scale. This
idea has been exploited in the seesaw mechanism to explain the smallness of
the neutrino masses relative to the
charged-lepton masses \cite{seesaw}.
Inspired by the indication for non-zero neutrino masses from neutrino
oscillations, model building
for incorporating neutrino masses and lepton mixing
and for explaining features of the neutrino mass spectrum
and of the mixing
matrix is in full swing. In this context, the seesaw mechanism
plays an important role \cite{barr}.

Usually, the lowest order in the inverse heavy seesaw scale is sufficient for
the purpose of
describing the light-neutrino mass matrix. However, one can
envisage situations which warrant going beyond the lowest order, like 
\begin{itemize}
\item
the need to know the neutrino masses to a better precision;
\item
a degeneracy
of the mass spectrum at lowest order, which is lifted at higher orders;
\item
the presence of massless neutrinos at lowest order,
which become massive at higher orders;
\item
a lepton mixing matrix with
some features at lowest order,
which disappear at higher orders \cite{silva};
\item
light mass terms which,
though much smaller than the heavy ones,
are not all of the same order of magnitude;
or,
inversely,
heavy masses with different orders of magnitude \cite{we};
or the same situation for both types of mass terms. 
\end{itemize}
In this letter we propose a
simple, minimal
procedure which
allows one to
decouple the small seesaw scale from the large one
order by order in the
inverse large scale. The procedure is defined
by a particular {\it Ansatz\/}
for the unitary matrix $W$ which performs this
decoupling, together with
a series expansion for $W$
in the inverse large scale. This procedure
enables, in principle, the calculation
of $W$, of the mass matrix of the 
light neutrinos, and of the mass matrix of the
heavy neutrinos, up to any
arbitrarily large order in the inverse heavy mass scale.

Seesaw-type mass matrices,
with two different mass scales,
are not confined to neutrino physics.
In this letter we also show
how one may recursively
decouple the light from the heavy quarks
when one adds vector-like quarks to the Standard Model
and one assumes the gauge-invariant
mass terms to be very large.

In Section \ref{neu}
we develop the decoupling procedure in the
case of the usual seesaw mechanism for neutrinos, whereas in Section
\ref{vector} we consider the extension of the Standard Model with
vector-like quarks. In Section \ref{concl} we present our
conclusions. 

\section{Neutrinos}\label{neu}

Let us consider a general model in which,
to the $n_L$ generations of leptons of the Standard Model,
one adds $n_R$ right-handed
(singlet) neutrinos $\nu_R$ \cite{schechter}.
One may take the charge conjugates of the latter fields,
$\nu_R^c \equiv C \overline{\nu_R}^T$,
which are left-handed,
and put them together with the doublet left-handed neutrinos.
(The matrix $C$ is the Dirac--Pauli charge-conjugation matrix.)
The most general mass terms
for all the neutrinos fields are then
\begin{equation}
{\cal L}_{\rm mass} = {\textstyle \frac{1}{2}}
\left( \begin{array}{cc} \nu_L^T, & \left( \nu_R^c \right)^T
\end{array} \right)
\left( \begin{array}{cc} M_L & M_D^T \\ M_D & M_R \end{array} \right) C^{-1}
\left( \begin{array}{c} \nu_L \\ \nu_R^c \end{array} \right)
+ {\rm H.c.}
\label{neutrino masses}
\end{equation}
The $n_L \times n_L$ matrix $M_L$
and the $n_R \times n_R$ matrix $M_R$ are symmetric;
the matrix $M_D$ is $n_R \times n_L$ and arbitrary.

The mass terms in $M_D$ are $\left| \Delta I \right| = 1/2$, where $I$
denotes the weak isospin,
and originate, usually,
in the Yukawa couplings of the leptons to scalar doublets
which acquire a vacuum expectation value (VEV).
The mass terms in $M_L$ are $\left| \Delta I \right| = 1$ and are present
if there exists in the theory
either a fundamental or an effective Higgs triplet
with a VEV.
The mass terms in $M_R$ are gauge-invariant
and we shall assume them to be
much larger than the ones in either $M_D$ or $M_L$.
We shall moreover assume $M_R$ to be non-singular.
The order of magnitude of the eigenvalues of $\sqrt{M_R^\dagger M_R}$
will be denoted by $m_R$. We
shall assume that all matrix elements of $M_L$
and of $M_D$ are much smaller than $m_R$.
Under these assumptions,
there are in the theory
$n_R$ heavy neutrinos with mass of order $m_R$,
and $n_L$ light neutrinos
with masses which are, in the case $M_L = 0$,
suppressed by one or more inverse powers of $m_R$
\cite{seesaw}.\footnote{A ``singular seesaw
mechanism'' \cite{singular}, where the matrix $M_R$ is singular,
is also possible. If
$\sqrt{M_R^\dagger M_R}$ has $n_0$ eigenvalues 0, then we can define
$n'_L = n_L + n_0$
and $n'_R = n_R - n_0$,
and this singular seesaw case in contained mathematically
in the usual seesaw case with $n'_L$
left-handed doublets and $n'_R$ right-handed singlets. This is most
easily seen by going into a basis where $M_R$ is diagonal, with the
$n_0$ diagonal zero entries coming first.}
Note, however, that the notion ``order of magnitude $m_R$'' should not be
taken too literally. To allow for a more general situation 
it should be understood in the sense that
the eigenvalues of $\sqrt{M_R^\dagger M_R}$,
though not necessarily all of the same order of magnitude,
are all much larger than the
matrix elements of $M_L$ and $M_D$; the latter, too, may have
orders of magnitude different among themselves.

It is our aim to decouple the heavy from the light neutrino fields,
and to derive the effective mass matrices for each of them.
This is done by performing a unitary transformation of the neutrino fields
by means of an
$\left( n_L + n_R \right) \times \left( n_L + n_R \right)$ unitary matrix $W$,
\begin{equation}
\left( \begin{array}{c} \nu_L \\ \nu_R^c \end{array} \right) = W
\left( \begin{array}{c} \nu_{\rm light} \\ \nu_{\rm heavy} \end{array}
\right)_L,
\label{W transformation}
\end{equation}
and by requiring from this unitary transformation that
\begin{equation}
W^T \left( \begin{array}{cc} M_L & M_D^T \\ M_D & M_R \end{array} \right) W 
= \left( \begin{array}{cc} M_{\rm light} & 0 \\
0 & M_{\rm heavy} \end{array} \right),
\label{decoupling}
\end{equation}
where $M_{\rm light}$ and $M_{\rm heavy}$ are,
respectively,
$n_L \times n_L$ and $n_R \times n_R$ symmetric matrices.

In Eq.~(\ref{decoupling}) we require that
the transformation performed by the matrix $W$
has the property of leading to a zero $n_L \times n_R$
submatrix of the mass matrix.
On these general grounds,
$W$ must have $n_L n_R$ degrees of freedom,
and any degrees of freedom in $W$ beyond these ones
are not needed
and might even render its computation unnecessarily complicated.
We shall therefore make
the following {\it Ansatz\/}\footnote{In
Ref.~\cite{valle} an equivalent {\it Ansatz\/} has been proposed;
Ref.~\cite{pilaftsis} uses the {\it Ansatz\/} of Eq.~(\ref{W}) in a more
restrictive context.} for $W$:
\begin{equation}
W = \left( \begin{array}{cc}
\sqrt{1 - B B^\dagger} & B \\ - B^\dagger & \sqrt{1 - B^\dagger B}
\end{array} \right),
\label{W}
\end{equation}
where $B$ is an $n_L \times n_R$ matrix
which must be fixed as a function of $M_L$,
$M_D$,
and $M_R$.
The square roots in Eq.~(\ref{W}) should be understood as power series,
for instance,
\begin{equation}
\sqrt{1 - B B^\dagger} = 1
- {\textstyle \frac{1}{2}} B B^\dagger
- {\textstyle \frac{1}{8}} B B^\dagger B B^\dagger
- {\textstyle \frac{1}{16}} B B^\dagger B B^\dagger B B^\dagger
- \ldots
\label{power series}
\end{equation}
The matrix $W$ in Eq.~(\ref{W}) is unitary by construction.
Indeed,
Eq.~(\ref{W}) might be seen as a generalization,
for matrices,
of the usual form of a $2 \times 2$ orthogonal matrix
\[
\left( \begin{array}{cc} \sqrt{1 - \sin^2{\theta}} & \sin{\theta} \\
- \sin{\theta} & \sqrt{1 - \sin^2{\theta}} \end{array} \right).
\]

With the {\it Ansatz\/} for $W$ in Eq.~(\ref{W}),
the condition of the vanishing of the off-diagonal submatrices
in Eq.~(\ref{decoupling}) reads
\begin{equation}
B^T M_L \sqrt{1 - B B^\dagger}
+ \sqrt{1 - B^T B^\ast} M_D \sqrt{1 - B B^\dagger}
- B^T M_D^T B^\dagger
- \sqrt{1 - B^T B^\ast} M_R B^\dagger = 0.
\label{condition}
\end{equation}
This equation may be solved by assuming that $B$ is a power series in $1/m_R$,
namely,
\begin{eqnarray}
B &=& B_1 + B_2 + B_3 + B_4 + \ldots,
\label{B} \\
\sqrt{1 - B B^\dagger} &=& 1
- {\textstyle \frac{1}{2}} B_1 B_1^\dagger
- {\textstyle \frac{1}{2}} \left( B_1 B_2^\dagger + B_2 B_1^\dagger \right)
\nonumber \\ & &
- {\textstyle \frac{1}{2}}
\left( B_1 B_3^\dagger + B_2 B_2^\dagger + B_3 B_1^\dagger
+ {\textstyle \frac{1}{4}} B_1 B_1^\dagger B_1 B_1^\dagger \right)
- \ldots,
\label{X}
\end{eqnarray}
where $B_j$ is by definition proportional to $\left( m_R \right)^{-j}$.
The recursive solubility of Eq.~(\ref{condition}) can be seen in the
following way. At order $\left( m_R \right)^0$
Eq.~(\ref{condition}) simply reads
$M_D - M_R B_1^\dagger = 0$, and this fixes $B_1$.
We then note that 
the $\ell$-th order of $\sqrt{1 - B B^\dagger}$
is a function of the
$B_j$ with $j<\ell$, {\it cf}.\ Eq.~(\ref{X}).
Following this observation, an inspection of the
$k$-th order of Eq.~(\ref{condition}) shows that, except for
a term $-M_R B_{k+1}^\dagger$,
all terms only involve the $B_j$ with $j \le k$.
In other words,
the $k$-th order terms of Eq.~(\ref{condition})
lead to an expression for $B_{k+1}$
in terms of the $B_j$ with $j \le k$.
This proves the recursive calculability of $B$.

From Eq.~(\ref{condition}) one thus obtains
\begin{eqnarray}
M_R B_1^\dagger &=& M_D,
\label{B1} \\
M_R B_2^\dagger &=& {M_R^{-1}}^\ast M_D^\ast M_L,
\label{B2} \\
M_R B_3^\dagger &=&
{M_R^{-1}}^\ast M_R^{-1} M_D M_L^\ast M_L 
- {M_R^{-1}}^\ast M_D^\ast M_D^T M_R^{-1} M_D
- {\textstyle \frac{1}{2}} M_D M_D^\dagger {M_R^{-1}}^\ast M_R^{-1} 
M_D.
\hspace*{9mm}
\label{B3}
\end{eqnarray}
The expressions for the $B_k$ with $k \ge 4$ 
are rather long and complicated,
but no fundamental problem arises in their recursive computation.
The effective mass matrices for the light and heavy neutrinos are then
\begin{eqnarray}
M_{\rm light} &=& M_L - M_D^T M_R^{-1} M_D
- {\textstyle \frac{1}{2}} \left( M_D^T M_R^{-1} {M_R^{-1}}^\ast M_D^\ast M_L
+ M_L M_D^\dagger {M_R^{-1}}^\ast M_R^{-1} M_D \right)
\nonumber \\ & &
+ {\textstyle \frac{1}{2}} \left[
M_D^T M_R^{-1} \left( M_D M_D^\dagger {M_R^{-1}}^\ast
+ {M_R^{-1}}^\ast M_D^\ast M_D^T \right) M_R^{-1} M_D \right.
\nonumber \\ & &
\left. - M_D^T M_R^{-1} {M_R^{-1}}^\ast M_R^{-1} M_D M_L^\ast M_L
- M_L M_L^\ast M_D^T M_R^{-1} {M_R^{-1}}^\ast M_R^{-1} M_D
\right] + \ldots, \hspace*{8mm}
\label{Mlight} \\
M_{\rm heavy} &=& M_R + {\textstyle \frac{1}{2}}
\left( M_D M_D^\dagger {M_R^{-1}}^\ast
+ {M_R^{-1}}^\ast M_D^\ast M_D^T \right)
\nonumber \\ & & + {\textstyle \frac{1}{2}}
\left( M_D M_L^\ast M_D^T M_R^{-1} {M_R^{-1}}^\ast
+ {M_R^{-1}}^\ast M_R^{-1} M_D M_L^\ast M_D^T \right) + \ldots,
\label{Mheavy}
\end{eqnarray}
respectively.
Equations~(\ref{Mlight}) and (\ref{Mheavy}) are correct up to order 
$\left( m_R \right)^{-3}$ and $\left( m_R \right)^{-2}$, respectively.

In the most common case of the seesaw mechanism,
which is for instance obtained by
extending the Standard Model with right-handed neutrino singlets
but avoiding the presence of any Higgs triplet, 
one has a vanishing matrix $M_L$. Then the recursive decoupling
mechanism becomes much simpler because
\begin{equation}\label{ML=0}
M_L = 0 \quad \Rightarrow \quad
B_2 = B_4 = B_6 = \ldots = 0.
\end{equation}
Let us sketch a proof of Eq.~(\ref{ML=0}) by induction. By direct
calculation we find that
$B_2 = 0$ when $M_L = 0$, see Eq.~(\ref{B2}).
Now we assume
that $B_2 = B_4 = \ldots = B_{2\ell} = 0$ ($\ell \ge 1$). In order to
calculate $B_{2\ell + 2}$ from Eq.~(\ref{condition}), we need to know
all the $B_j$ with $j \le 2\ell + 1$. Then, according to the induction
assumption, only the $B_j$ with odd index $j$ contribute to 
$B_{2\ell + 2}$. We now notice that $\sqrt{1-BB^\dagger}$ only has even
powers of $1/m_R$ up to order $2\ell + 1$. Then, a term-by-term inspection
of Eq.~(\ref{condition}) with $M_L=0$ reveals that all terms are zero at order
$2\ell + 1$ except for the term $-M_R B_{2\ell + 2}^\dagger$.
Therefore, $B_{2\ell + 2} = 0$, and this proves the theorem
in Eq.~(\ref{ML=0}).
As a consequence, we have
\begin{eqnarray}
B &=& B_1 + B_3 + \ldots,
\\
\sqrt{1 - B B^\dagger} &=& 1
- {\textstyle \frac{1}{2}} B_1 B_1^\dagger
- {\textstyle \frac{1}{2}}
\left( B_1 B_3^\dagger+ B_3 B_1^\dagger
+ {\textstyle \frac{1}{4}} B_1 B_1^\dagger B_1 B_1^\dagger \right)
- \ldots,
\end{eqnarray}
with
\begin{eqnarray}
M_R B_1^\dagger &=& M_D,
\\
M_R B_3^\dagger &=& - {M_R^{-1}}^\ast M_D^\ast M_D^T M_R^{-1} M_D
- {\textstyle \frac{1}{2}} M_D M_D^\dagger {M_R^{-1}}^\ast M_R^{-1} M_D,
\end{eqnarray}
and, using similar considerations as in the above proof by induction,
we also find that only odd powers of $1/m_R$
are present in $M_{\rm light}$ and $M_{\rm heavy}$, in particular,
\begin{eqnarray}
M_{\rm light} &=& - M_D^T M_R^{-1} M_D
\nonumber \\ & &
+ {\textstyle \frac{1}{2}}
M_D^T M_R^{-1} \left( M_D M_D^\dagger {M_R^{-1}}^\ast
+ {M_R^{-1}}^\ast M_D^\ast M_D^T \right) M_R^{-1} M_D + \ldots,
\label{Ml} \\
M_{\rm heavy} &=& M_R + {\textstyle \frac{1}{2}}
\left( M_D M_D^\dagger {M_R^{-1}}^\ast
+ {M_R^{-1}}^\ast M_D^\ast M_D^T \right) + \ldots
\end{eqnarray}
It is possible to proceed and derive,
recursively,
expressions for $B_5$,
$B_7$,
and so on,
and also better approximations
for $M_{\rm light}$ and for $M_{\rm heavy}$.
However,
that task is tedious and complicated,
since the number of terms in the polynomials increases very rapidly.

The subcase of the case $M_L = 0$ where the leading term in 
$M_{\rm light}$ (\ref{Ml}) is zero has been studied in
Ref.~\cite{pilaftsis}. In this subcase one does not need recursive
expansions, as one is able to give closed forms for all the relevant 
matrices. By using Eq.~(\ref{condition}) and the definition 
$Z = M_R^{-1} M_D$, one readily verifies that \cite{pilaftsis}
\begin{equation}
M_L = 0\,, \quad M_D^T M_R^{-1} M_D = 0 \quad \Rightarrow \quad
B = Z^\dagger \left( 1 + Z Z^\dagger \right)^{-1/2} =  
\left( 1 + Z^\dagger Z \right)^{-1/2} Z^\dagger\,.
\end{equation}
Then, using Eq.~(\ref{decoupling}), one finds after some algebra that
\begin{equation}
M_{\rm light} = 0 \quad \mbox{and} \quad
M_{\rm heavy} = \left( 1 + Z^* Z^T \right)^{1/2} M_R
\left( 1 + Z Z^\dagger \right)^{1/2}.
\end{equation}
It is remarkable that the vanishing of the lowest order
contribution to $M_{\rm light}$ in Eq.~(\ref{Ml})
entails the exact vanishing of the whole matrix.\footnote{We thank
A.\ Pilaftsis and A.\ Joshipura for calling our attention to this
interesting subcase.} 

\section{Vector-like quarks}\label{vector}

Let us consider that to the $n_g$ generations
of $u$ and $d$ quarks of the Standard Model
one adds $n^\prime$ vector-like singlet quarks of,
say,
charge $-1/3$.
One then has $n_g$ left-handed quarks $d_L$,
which are components of doublets of SU(2),
together with $n^\prime$ left-handed quarks $D_L$,
which are singlets of SU(2);
the $n_g + n^\prime$ right-handed quarks $d_R$ are all singlets \cite{book}.
The mass term for these down-type quarks reads
\begin{equation}
\left( \begin{array}{cc} \bar d_L, & \bar D_L \end{array} \right)
{\cal M}\, d_R + {\rm H.c.},\ {\rm with}\ {\cal M} =
\left( \begin{array}{c} m \\ M \end{array} \right).
\end{equation}
The matrix $m$ is $n_g \times \left( n_g + n^\prime \right) $
and gives the $\left| \Delta I \right| = 1/2$ mass terms;
the matrix $M$ is $n^\prime \times \left( n_g + n^\prime \right) $
and contains the SU(2)-invariant mass terms.
We shall assume that the matrix elements of $m$ are of order
$\lambda$, and that the $n'$ eigenvalues of $\sqrt{MM^\dagger}$
are all non-zero and
of order $\Lambda$, 
with $\lambda / \Lambda \equiv \epsilon \ll 1$.

The squared-mass matrix for the left-handed quarks $d_L$ and $D_L$ is
\begin{equation}
{\cal M} {\cal M}^\dagger =
\left( \begin{array}{cc} m m^\dagger & m M^\dagger \\
M m^\dagger & M M^\dagger \end{array} \right).
\end{equation}
After diagonalization,
there will be $n_g$ ``light'' quarks with masses of order $\lambda$,
and $n^\prime$ ``heavy'' quarks with masses of order $\Lambda$.
One would like to decouple the light quarks from the heavy
ones and to find
separate squared-mass matrices for each category.
This we shall do by means of a unitary transformation,
\begin{equation}
U^\dagger {\cal M} {\cal M}^\dagger U =
\left( \begin{array}{cc} D_{\rm light} & 0 \\ 0 & D_{\rm heavy} 
\end{array} \right),
\label{decoupling1}
\end{equation}
where $U$ is $\left( n_g + n^\prime \right)
\times \left( n_g + n^\prime \right)$ unitary,
while $D_{\rm light}$ and $D_{\rm heavy}$ are,
respectively,
$n_g \times n_g$ and $n^\prime \times n^\prime$ Hermitian
(and positive definite) matrices.

In Eq.~(\ref{decoupling1}) we are,
as a matter of fact,
transforming an $n_g \times n^\prime$ submatrix
of ${\cal M} {\cal M}^\dagger$ to zero by means of a unitary transformation.
It is therefore convenient to choose for the unitary matrix $U$
of that transformation an ${\it Ansatz}\/$ which has exactly
$n_g n^\prime$ (complex) degrees of freedom.
We therefore write
\begin{equation}
U = \left( \begin{array}{cc}
\sqrt{1 - F F^\dagger} & F \\ - F^\dagger & \sqrt{1 - F^\dagger F}
\end{array} \right),
\label{form of U}
\end{equation}
where $F$ is an arbitrary $n_g \times n^\prime$ matrix,
to be fixed as a function of $m$ and of $M$.
The square roots in Eq.~(\ref{form of U}) should be understood as power series,
just as in Eq.~(\ref{power series}).
It is then obvious that the matrix $U$ in Eq.~(\ref{form of U})
is indeed unitary;
it is the analogue of the matrix $W$ in the
previous section. 

The condition which fixes $F$ is the demand that
there is a zero submatrix
in Eq.~(\ref{decoupling1}),
namely,
\begin{equation}
\sqrt{1 - F F^\dagger}\, m m^\dagger F
+ \sqrt{1 - F F^\dagger}\, m M^\dagger \sqrt{1 - F^\dagger F}
- F M m^\dagger F
- F M M^\dagger \sqrt{1 - F^\dagger F} = 0.
\label{condition 2}
\end{equation}
We solve this equation by writing $F$ as a power series in $\epsilon$,
\begin{equation}
F = F_1 + F_3 + F_5 + \ldots,
\end{equation}
where $F_j$ has matrix elements of order $\epsilon^j$.
Notice that only terms with odd $j$ are necessary.
Defining
\begin{equation}
S \equiv \left( M M^\dagger \right)^{-1} = S^\dagger,
\label{S definition}
\end{equation}
it is easy to find
\begin{eqnarray}
F_1 &=& m M^\dagger S,
\label{F1} \\
F_3 &=& m \left( m^\dagger m M^\dagger S
- M^\dagger S M m^\dagger m M^\dagger S
- {\textstyle \frac{1}{2}} M^\dagger S^2 M m^\dagger m M^\dagger
\right) S.
\label{F3}
\end{eqnarray}
Unfortunately,
the expressions for the $F_j$ become quite complicated when $j \ge 5$,
although they may in principle be derived recursively
without any problem.

One is thus able to decouple
the light from the heavy quarks.
One finds their respective squared-mass matrices to be given by
\begin{eqnarray}
D_{\rm light} &=& m \left( 1 - M^\dagger S M \right) m^\dagger
\nonumber \\
  & &
- {\textstyle \frac{1}{2}} m \left[
\left( 1 - M^\dagger S M \right) m^\dagger m M^\dagger S^2 M
+ M^\dagger S^2 M m^\dagger m \left( 1 - M^\dagger S M \right)
\right] m^\dagger
\nonumber \\ && + \dots \label{d} \\
D_{\rm heavy} &=& M M^\dagger
+ {\textstyle \frac{1}{2}} \left( S M m^\dagger m M^\dagger
+ M m^\dagger m M^\dagger S \right) + \ldots
\label{D}
\end{eqnarray}
The expressions of $D_{\rm light}$ and $D_{\rm heavy}$ have been given
to subleading order in $\epsilon^2$.
Note that only even powers of $\epsilon$
appear in the expansions of these squared-mass matrices.

It is interesting to write down the equations above
in a specific simple weak basis.
By performing a unitary mixing of the right-handed singlet quarks $d_R$
and also of the left-handed singlet quarks $D_L$,
one finds a weak basis in which
\begin{equation}
m = \left( \begin{array}{cc} G, & J \end{array} \right)\
{\rm and}\
M = \left( \begin{array}{cc} 0, & \hat M \end{array} \right),
\label{weak basis}
\end{equation}
where $\hat M$ is a diagonal and positive
$n^\prime \times n^\prime$ matrix,
while $G$ and $J$ are general
$n_g \times n_g$ and $n_g \times n^\prime$ matrices,
respectively.
In this weak basis one finds
\begin{eqnarray}
D_{\rm light} &=& G G^\dagger - {\textstyle \frac{1}{2}}
\left\{ G G^\dagger, J \hat M^{-2} J^\dagger \right\} + \ldots,
\label{d weak basis} \\
D_{\rm heavy} &=& \hat M^2 + {\textstyle \frac{1}{2}}
\left( \hat M^{-1} J^\dagger J \hat M + 
\hat M J^\dagger J \hat M^{-1} \right) + \ldots,
\label{D weak basis}
\end{eqnarray}
where $\left\{ X, Y \right\} \equiv X Y + Y X$ 
is the anticommutator.
In this basis it is also obvious that the
$(n_g+n') \times (n_g+n')$
matrix $M^\dagger S M$, which appears in Eq.~(\ref{d}), is a projector,
which projects into
the space characterized by the last $n'$ indices. 
Therefore, at lowest order,
in the matrix $D_{\rm light}$ of Eq.~(\ref{d})
only the left $n_g \times n_g$ submatrix
of $m$ contributes.
The degrees of freedom beyond the
Standard Model, which are characterized in the weak basis of
Eq.~(\ref{weak basis}) by the last $n'$ indices, do not contribute to
the light-quark masses, as expected.

\section{Conclusions}\label{concl}

In this note we have discussed a method which allows to separate,
order by order in the inverse seesaw scale,
the small and large scales in seesaw-type mass matrices,
and to obtain a recursive expansion for the light-neutrino
and for the heavy-neutrino mass matrices.
This has been achieved by the {\it
Ansatz\/} of Eq.~(\ref{W}) for the matrix $W$, 
which approaches the unit matrix
when the large scale tends to infinity.
A similar method,
with the matrix $U$ of Eq.~(\ref{form of U}),
can be used to decouple the light quarks
from additional heavy quarks in extensions of the Standard Model
with vector-like quarks.
The lepton mixing matrix and the quark mixing matrix
can then be calculated from the matrices $M_{\rm light}$ of Eq.~(\ref{Mlight})
and $D_{\rm light}$
of Eq.~(\ref{d}), respectively.
For this purpose, the lowest order of these matrices might be sufficient,
yet in some
instances it might be necessary to include further terms
in order to obtain trustworthy representations of the mixing matrices.
If one wants to go beyond
lowest-order mixing, then in the neutrino case
there are contributions from higher orders in $M_{\rm light}$
and in $W$; the same holds in the vector-like-quark case
for $D_{\rm light}$ and $U$.

\end{document}